\def\flux{{\rm erg \, s^{-1} \, cm^{-2}}}
\def\sb{{\rm erg \, s^{-1} \, cm^{-2} \, arcmin^{-2}}}

\def\etal{{et al. }}
\def\lum{{\rm erg \, s^{-1}} }
\def\asca{{\em ASCA~ }}
\def\asca_kT{{$5.7^{+2.1}_{-1.1}$~ }}

\def\chisq{$\chi^2$}

\documentstyle[12pt,aaspp4,epsf]{article}
\begin{document}
\lefthead{Donahue}
\righthead{MS1137.5+6625}
\slugcomment{Accepted \apj, June 16, 1999}
\title{The Second Most 
Distant Cluster of Galaxies in the EMSS}
\author{Megan Donahue, G. Mark Voit, Caleb A. Scharf} 
\affil{Space Telescope Science Institute \\3700 San Martin Drive \\
Baltimore, MD 21218 \\ donahue@stsci.edu, voit@stsci.edu, scharf@stsci.edu}
\author{Isabella M. Gioia\altaffilmark{1,2}, Christopher R. Mullis
\altaffilmark{2}}
\altaffiltext{1}{Home Institution: Istituto di Radioastronomia del CNR, Via
Gobetti, 101 40129 Bologna, Italy}
\altaffiltext{2}{Visiting Astronomers at the W. M. Keck Observatory, jointly
operated by the California Institute for Technology, the University of 
California, and the National Aeronautics and Space Administration.}
\affil{Institute for Astronomy, 2680 Woodlawn Drive \\ Honolulu, HI 96822 
\\ gioia@galileo.ifa.hawaii.edu, mullis@ifa.hawaii.edu}
\author{John P. Hughes}
\affil{Department of Physics and Astronomy\\ Rutgers University \\ 136
Frelinghuysen Road \\ Piscataway, NJ 08854-8019 \\ jph@physics.rutgers.edu}
\author{John T. Stocke}
\affil{University of Colorado \\ Center for Astrophysics and Space
Astronomy \\ CB 389 \\ Boulder CO 80309 \\ 
stocke@casa.colorado.edu}

\begin{abstract}

We report on our ASCA, Keck, and ROSAT observations of MS1137.5+6625,
the second most distant cluster of galaxies in the Einstein Extended 
Medium Sensitivity Survey (EMSS), at redshift 0.78.  We now
have a full set of X-ray temperatures, optical velocity dispersions,
and X-ray images for a complete, high-redshift sample of clusters of
galaxies drawn from the EMSS.  Our ASCA observations of MS1137.5+6625
yield a temperature of \asca_kT keV and a metallicity of 
$0.43^{+0.40}_{-0.37}$ solar, with 90\% confidence limits. Keck-II
spectroscopy of 22 cluster members reveals a velocity dispersion
of $884^{+185}_{-124}$ km s$^{-1}$. 
This cluster is the most distant in the sample with a detected iron line. 
We also derive a mean abundance at $z=0.8$ by simultaneously fitting
X-ray data for the two $z=0.8$ clusters, and obtain an abundance of
$Z_{Fe} = 0.33 \pm^{0.26}_{0.23}$. 
Our ROSAT observations show that MS1137.5+6625 is regular and highly centrally concentrated.  Fitting of a $\beta$ model to the X-ray surface brightness
yields a core radius of only 71 $h^{-1}$ kpc ($q_0=0.1$) with 
$\beta=0.70 \pm^{0.45}_{0.15}$.  The gas mass interior to 0.5 h$^{-1}$ Mpc 
is thus $1.2 \pm^{0.2}_{0.3} \times 10^{13} h^{-5/2} M_\odot$ ($q_0=0.1$). 
If the cluster's gas is nearly isothermal and in hydrostatic 
equilibrium with the cluster potential, the total mass of the cluster
within this same region is $2.1 \pm ^{1.5}_{0.8}  \times 10^{14} h^{-1} 
M_\odot$, giving a gas fraction of $0.06\pm 0.04 ~ h^{-3/2}$. This cluster
is the highest redshift EMSS cluster showing 
evidence for a possible cooling flow
($\sim 20-400$ M$_\odot$ yr$^{-1}$). 
The velocity dispersion, temperature, gas fraction, and
iron abundance of MS1137.5+6625 are all statistically the same as 
those properties
in lower redshift clusters of similar luminosity.
With this cluster's temperature now in hand, we derive a high-redshift 
temperature function for EMSS clusters at $0.5 < z < 0.9$ and compare 
it with temperature functions at lower redshifts, showing that the
evolution of the temperature function is relatively modest.
Supplementing our high-redshift sample with other data from the 
literature, we demonstrate that neither the cluster luminosity-temperature 
relation, nor cluster metallicities, nor the cluster gas fraction has 
detectably evolved with redshift. The very modest 
degree of evolution in the luminosity-temperature relation inferred from
these data is inconsistent with the 
absence of evolution in the X-ray luminosity functions derived from ROSAT 
cluster surveys if a critical-density structure formation model is
assumed.

\end{abstract}


\section{Introduction}

The Extended Medium Sensitivity Survey (EMSS) sample of high-redshift 
clusters of galaxies serendipitously discovered in the fields of 
{\em Einstein} Imaging Proportional Counter (IPC) images (Gioia \etal
1990; Henry \etal 1992) has  
proved to be useful for testing cosmological models (Henry 1997;
Eke \etal 1998; Donahue \etal 1998). It was the first X-ray survey with significant numbers
of clusters at $z > 0.3$.  Even now, the EMSS stands unique among
cluster surveys.  Because of its large sky coverage and moderately
deep X-ray sensitivity, it is the only survey with a full suite of 
spectroscopic redshifts that can begin to place
constraints on the evolution of the rarest and most luminous clusters
of galaxies. These are the clusters which are expected to evolve the
most dramatically in models of structure formation driven by gravitational
collapse. The only survey similar in size and depth is the ROSAT serendipitous 
survey by Vikhlinin \etal (1998), containing 200 galaxy clusters with
a mixture of photometric and spectroscopic redshifts. 

 The clusters we have studied from the EMSS are at moderately
high redshift ($z>0.5$) and high X-ray luminosities ($\sim 10^{45}~ \lum$), 
and thus are the clusters most likely to show evidence for evolution, if
cluster evolution occurs. The original luminosity functions derived from the 
EMSS suggested that the highest luminosity clusters $L_x \gtrsim 
10^{44} h^{-2}~ \lum$ may be somewhat less
common at the highest redshifts (Gioia \etal 1990; Henry \etal 1992). More
recently, deeper X-ray surveys with more sophisticated cluster detection
algorithms but less sky coverage (e.g., Rosati \etal 1998; Jones \etal 1998)
showed little evolution for cluster luminosities $\lesssim 8 \times 10^{43}~ h^{-2} \lum$.
 Followup X-ray observations of the EMSS clusters 
with ASCA and ROSAT to acquire temperatures and emitting volumes of
the intracluster gas  
(Donahue 1996; Donahue \etal 1998; Furuzawa \etal 1994; Yamashita 
\etal 1994) along with ground-based measurements of their galaxy velocity 
dispersion (Carlberg \etal 1996; Donahue 1996; Donahue \etal 1998) 
have shown that these clusters contain hot intracluster 
media and correspondingly high velocity dispersions characteristic of 
massive clusters.  In addition, 
the weak lensing signatures of these clusters corroborate the large mass 
estimates derived from their X-ray temperatures and galaxy kinematics 
(Luppino \& Gioia 1995; Luppino \& Kaiser 1996).

The temperature function of massive distant clusters strongly constrains
cosmological models because cluster temperatures are closely related
to cluster masses whose evolution with redshift is quite sensitive
to cosmological parameters.  Gravitational compression is the dominant 
source of heating of the intracluster medium (ICM) of massive clusters. 
While the ICM of smaller clusters ($\lesssim 2$~keV) may be significantly 
modified by the effects of superwinds or other energetic processes,  
the gravitational energy per particle in larger clusters is several 
times the thermal energy per particle available through supernovae 
(e.g., Renzini \etal 1993).  
Because the temperatures of clusters reflect their virial
masses, a compilation of the cluster temperature function at different 
epochs reflects how the cluster mass function evolves.  Furthermore the 
evolution of the cluster mass function is exponentially sensitive 
to the mean density of the universe, so even a handful of massive clusters
at high redshift can begin to constrain $\Omega_M$, the fraction of
the critical density in the form of gravitating matter (Eke \etal 1998,
Donahue \etal 1998, Bahcall, Fan \& Cen 1997; Viana \& Liddle 1996; 
Oukbir \& Blanchard 1992; Eke, Cole \& Frenk 1996; Bahcall \& Fan 1998).

This paper reports on our observations of the complete EMSS sample
of high-redshift clusters.  Section~2 presents our observations of 
the cluster MS1137.5+6625, the last of the high-redshift EMSS clusters 
to be observed with ASCA, the ROSAT HRI, and Keck-II.  Section~3 summarizes
and updates our findings for the sample as a whole and presents
the temperature function for EMSS clusters at $z > 0.5$, the first
derived from a complete sample at such high redshifts.  Section~4
assesses whether the X-ray luminosity-temperature relation for clusters
evolves with redshift and finds no evidence for significant evolution.
Section~5 summarizes our results. In this paper we parametrize 
$H_0$ as $100 \, h \, {\rm km \, s^{-1} \, Mpc^{-1}}$, and 
explicitly state the $q_0$ assumed.

\section{Cluster MS1137.5+6625}

Cluster MS1137.5+6625 at $z = 0.78$ is the last of the high-redshift 
EMSS clusters to be observed in our program.  Our ASCA
observations find a best-fit rest-frame temperature for this
cluster of \asca_kT~keV with a best-fit iron abundance
of $0.43^{+0.40}_{-0.37}$ times solar (90\% confidence intervals).
The ROSAT observations indicate that the cluster is relatively
compact, with a core radius of roughly $60-70 \, h^{-1} \, 
{\rm kpc}$.  The velocity dispersion for the cluster derived from
Keck-II observations of 22 cluster members ($884^{+185}_{-124} \,
{\rm km \, s^{-1}}$) is consistent with the ASCA X-ray temperature.
Because the uncertainties on the derived iron abundance are so large, 
we simultaneously fit the ASCA data on MS1137.5+6625 and MS1054.4-0321
($z = 0.83$) while constraining their iron abundances to be the same in order
to derive a ``mean'' iron abundance for clusters at $z \sim 0.8$
equal to $0.33^{+0.26}_{-0.23}$ times solar (90\% confidence).
The remainder of this section provides the details of these
observations.

\subsection{ASCA Observations}

ASCA executed a single long ($\sim 70,000$ s) exposure
of the cluster MS1137.5+6625 during 1997 May 3-8.  
This satellite carries four independent X-ray telescopes,
two with Gas Imaging Spectrometers (GIS) and two with
Solid-State Imaging Spectrometers (SIS), from which we
obtained four independent datasets.
To prepare the data for analysis, we extracted clean 
X-ray event lists using a magnetic cut-off rigidity 
threshold of 4 GeV~c$^{-1}$ and the recommended minimum elevation 
angles and bright Earth angles to reject background contamination 
(see Day \etal 1995). 
Events were extracted for spectral analysis 
from within circular apertures of radii  
1.6 arcmin (SIS1), 2.3 arcmin (SIS0), and 3.25 arcmin (GIS) 
to maximize the ratio of signal to background noise for the
spectra analysis. The SIS apertures are smaller than usual 
because the cluster was not centered on the SIS detector. To
obtain the best estimates of the total flux from the GIS 
observations, we increased the radius of the GIS aperture to
5.0 arcmin; these data were used for flux estimates only, not
to obtain temperature estimates. 
We rejected $\sim$98\% of cosmic ray events by using only SIS 
chip data grades of 0, 2, 3, 4, and we rejected hot and 
flickering pixels. Light curves for each instrument were 
visually inspected, and time intervals with high background 
or data dropouts were excluded manually.
We rebinned the SIS data in the standard way to 512 spectral 
channels using Bright2Linear (see Day \etal 1995) with the lowest
13 channels flagged as bad, and we regrouped all the spectral 
data so 
that no energy bin under consideration 
had fewer than 25 counts for the GIS detector
and 16 counts for the SIS detector. 
Table~\ref{asca} gives the resulting net count rates
and effective exposure times. We do not expect the derived
X-ray fluxes to be consistent between the two SIS detectors 
and the GIS detectors because
the target was not centered and some of the extended flux may 
have missed the SIS detectors.

Background estimates were taken from the regions of the detector
surrounding the cluster observation. The cluster is very compact,
and local backgrounds have proved reliable for all of our previous observations
(Donahue 1996; Donahue \etal 1998). 
After rebinning the data, we restricted our fitting of the ASCA spectra
to the $0.5-8$~keV range over which the signal-to-noise was adequate after
background subtraction.

To analyze the spectra, we used XSPEC (v10.0) from the software 
package XANADU available from the ASCA GOF (Arnaud 1996). 
We fitted the spectral data from the four ASCA datasets and their 
respective response files (see Day \etal 1995).  The individual
SIS response matrices were generated with the tool {\em sisrmg} 
(1997 version), which takes into account temporal variations 
in the gain and removes the inconsistencies seen in data analyzed 
with the standard SIS response matrices.

\begin{figure}
\plotone{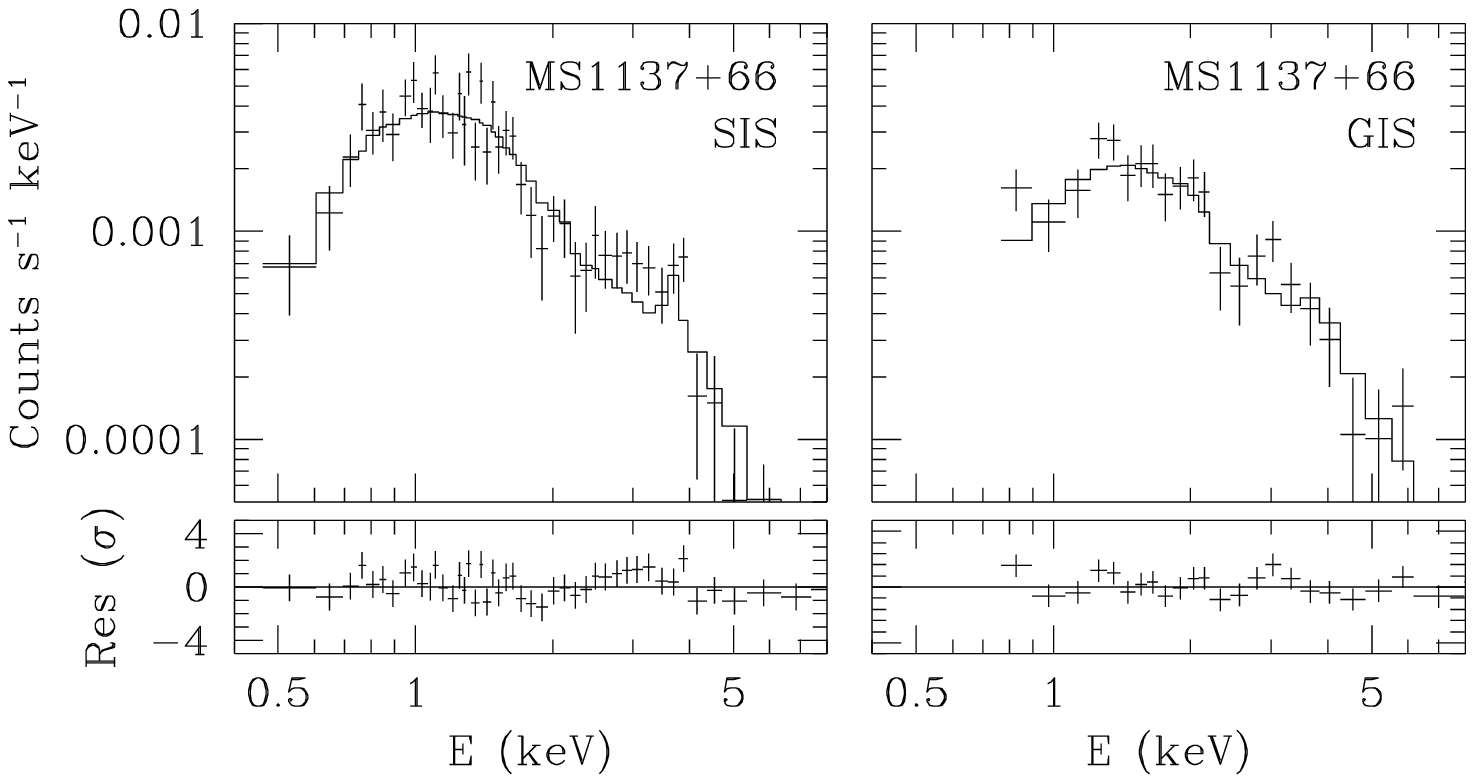}
\figcaption[ms1137_jph.ps]{Binned, background-subtracted 
ASCA spectra and the best-fit thermal model. For
clarity of display only, the individual SIS and GIS X-ray spectra and
corresponding models were averaged to produce this figure.\label{spectra}}
\end{figure}

The standard model we used to fit the data was a Raymond-Smith 
thermal plasma with a temperature $T_X$, absorbed by cold Galactic 
gas with a characteristic column density of neutral hydrogen $N_{\rm H}$.
The Galactic column density toward the cluster is not constrained by
the data.  Thus, we fixed soft X-ray absorption at an assumed Galactic HI value of
$1.0 \times 10^{20}$ cm$^{-2}$ (Gioia \etal 1990). Allowing the modelled
$N_{\rm H}$ intrinsic to the cluster to vary had little 
effect on the best-fit temperature ($\sim 0.1$ keV) and no significant
improvement of reduced $\chi^2$. Thus, in the following
results, we fixed the intrinsic column to be zero. 
Each SIS spectrum was fit with its own normalization; 
the normalizations of the GIS spectra were constrained to be the same.  
The parameters varied to provide a fit to 4 spectra were therefore
the 3 independent normalizations (1 for each SIS spectrum, 1 for 
both GIS spectra), a single emission-weighted temperature, 
and the metallicity (effectively the iron abundance of the cluster
gas). The binned 
spectra and best-fit model convolved with the appropriate detector
response matrix are found in Figure~\ref{spectra}. Note that 
this figure shows the average SIS and GIS spectra for display
purposes only;
to reiterate: all spectral analyses were carried out on the individual datasets.
The best-fit temperature was \asca_kT keV 
with a best-fit iron abundance of $0.43 \pm^{0.40}_{0.37}$ solar, 
where solar abundances are those of Anders \& Grevesse (1989). 
  All uncertainties quoted are the 90\% 
confidence levels for a single interesting parameter ($\Delta \chi^2 = 2.70$), 
and all fits have an acceptable reduced  \chisq~ of $\sim0.8-0.9$.

The cluster flux in the 2-10 keV observed band within the
GIS3 5 arcmin aperture is $(6.1 \pm 0.7) \times 10^{-3}$ ct s$^{-1}$, 
corresponding to $1.8 \times 10^{-13} \flux$,
and the cluster luminosity in the 2-10 keV rest band is 
$2.8 \times 10^{44} \, h^{-2} \, \lum$ for $q_0=0.0$ and 
$1.9 \times 10^{44} \, h^{-2} \, \lum$ for $q_0=0.5$. 
This luminosity is consistent with the luminosity one would
predict for a cluster of this temperature, given the low-redshift
$L_x - T_x$ relation (Edge \& Stewart 1991; David \etal 1993).

\subsection{ROSAT Observations}

\begin{figure}
\figcaption[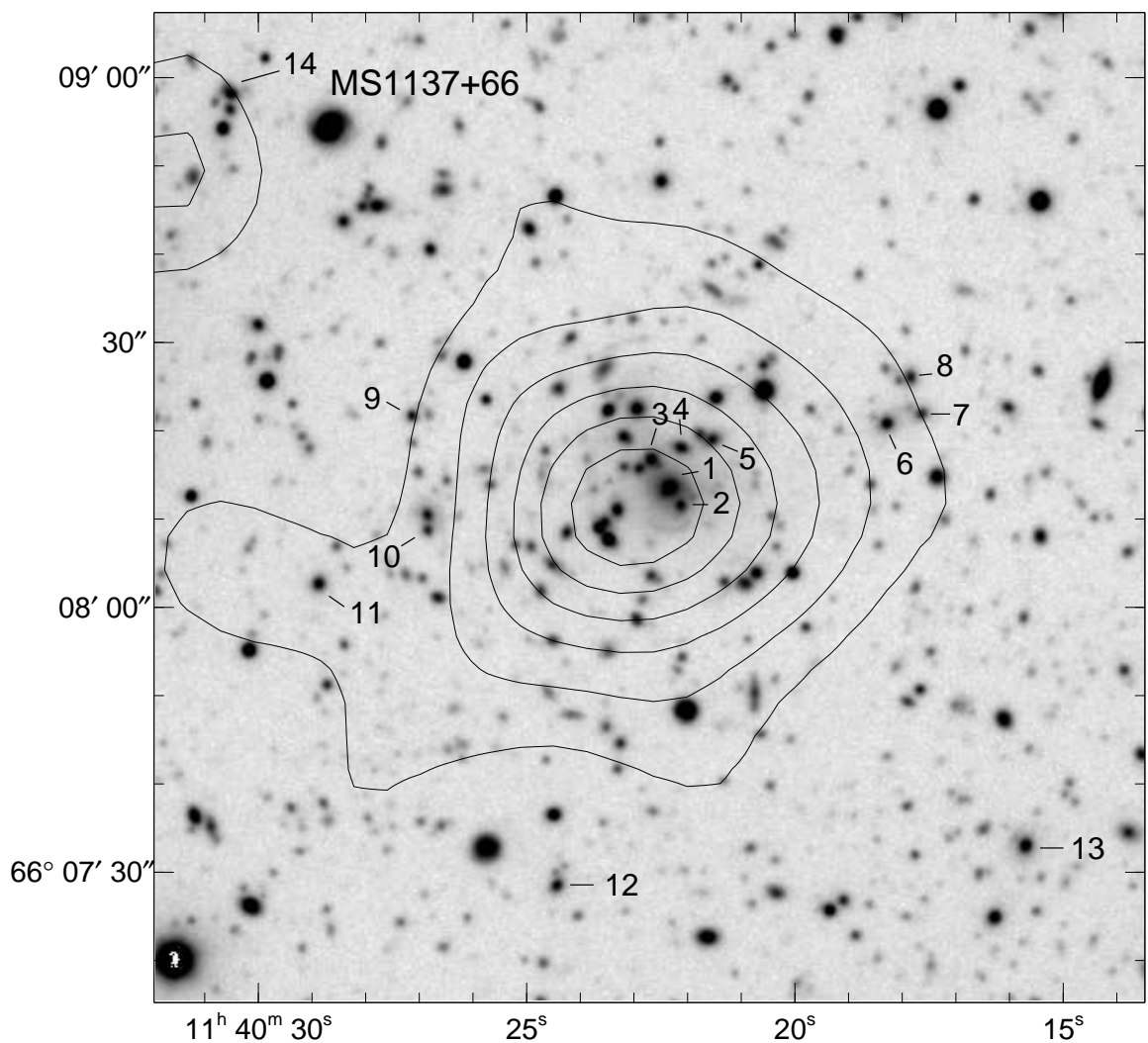]{An optical image of the central region of the 
X-ray cluster MS1137.5+6625, with X-ray emission contours overlaid. 
The ROSAT HRI image was adaptively smoothed with Gaussians such that
the product of the Gaussian widths and the number of counts inside
that Gaussian was nearly constant.  The contours start three 
sigma above the background ($5.2\times 10^{-3}$
cts s$^{-1}$ arcmin$^{-2}$) and increase in linear steps of  $1.57\times 10^{-3}$
cts s$^{-1}$ arcmin$^{-2}$.
The maximum contour value displayed is $1.30\times 10^{-2}$ cts s$^{-1}$ arcmin$^{-2}$.
The galaxies are identified according
to Table~\ref{galaxies}.\label{greyscale}}

\plotone{gioia_ms1137r_jack_small.eps}
\end{figure}

\begin{figure}
\plotone{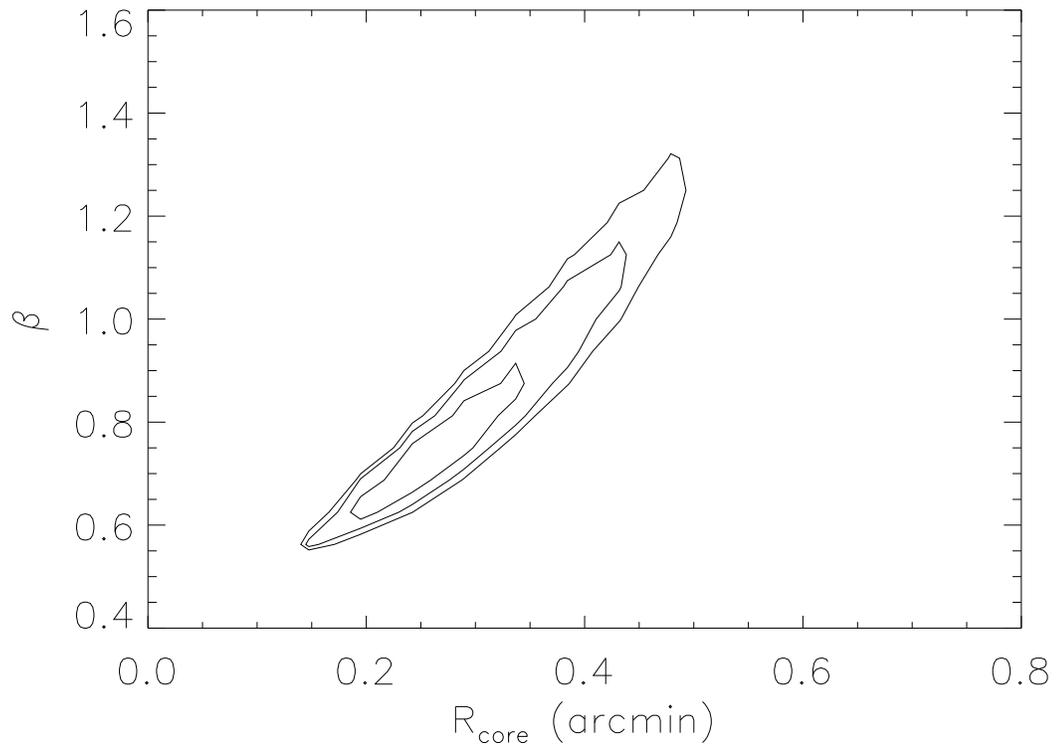}
\figcaption[ms1137_beta_radius_contour.ps]{Contour plot for the
two-dimensional uncertainties for $\beta$ and core radius for
MS1137, where $\Delta \chi^2 = 2.30$, 4.61, and 6.17 corresponding
to 68\%, 90\% and 95\% confidence intervals.\label{beta_core}}
\end{figure}

The ROSAT HRI obtained observations of MS1137.5+6625 on 3 separate
occasions for a total of 100,034 seconds (Table~\ref{hri}). 
The HRI data were filtered
to include only PHA bins 1-7. By excluding the higher PHA bins we
reduced the background count rate but sacrificed very few source counts.   
Contours of X-ray emission are overplotted on an optical image of
the cluster from Clowe et al. (1998) in Figure~\ref{greyscale}. 
We binned the HRI data into 4\arcsec\ by 4\arcsec\ bins and
fit a two-dimensional model to the X-ray emission by generating
an intrinsic model, convolving that model with the HRI point response
function, and fitting that to the binned data (see Hughes \&
Birkinshaw 1998 for a full description of the technique and
software). The background count rate in the HRI image
near the cluster was $3.59 \times 10^{-3}$ counts s$^{-1}$ arcmin$^{-1}$. 
The best fit model 
to the surface brightness was a circular King model 
($I \propto [1+(r/r_c)^2]^{-3\beta + 1/2}$) with a 
core radius $r_c$ of $0.25\arcmin \pm_{0.10}^{0.20}$ and
$\beta = 0.70 \pm_{0.15}^{0.45}$ (90\% confidence for
2 interesting parameters). In Figure~\ref{beta_core} we
plot the 68\%, 90\%, and 95\% 
confidence contours for two interesting parameters,
$\beta$ and core radius. The slope $\beta$ is not well
measured with the HRI data, but it is consistent with that of low
redshift clusters of galaxies. We also fit an elliptical King model
to the data, but without significant improvement in $\chi^2$. 
This cluster is more compact than many X-ray clusters, with a core
radius of only 71$h^{-1}$ kpc ($q_0=0.1$) to 61$h^{-1}$ kpc ($q_0=0.5$). 
The best-fit normalization corresponds to a central surface brightness
of $0.023$ HRI counts s$^{-1}$ arcmin$^{-2}$ which is comparable to
other high redshift clusters without strong cooling flows. MS0016+16,
for example, has a central surface brightness
of $0.047$ HRI counts s$^{-1}$ arcmin$^{-2}$ (Hughes \& Birkinshaw 1998).
The nominal center of the X-ray emission in MS1137.5+6625
is $11^h 40^m 22.0^s$, $+66^\circ \, 08\arcmin \, 17\arcsec$ (J2000).

We converted the observed central surface brightness to physical units
by assuming 1 HRI count = $3.77 \times 10^{-11}$ erg cm$^{-2}$. 
The calibration was derived by using the IRAF/PROS program called
{\em hxflux} to convert the count rate to an unabsorbed energy flux between
the ROSAT band energies of 0.2-2.5 keV, assuming
that $kT=5.7$~keV, log $N_H = 20.0$, $z=0.78$, and abundances
of 30\% solar. The conversion is not very sensitive to the assumed
abundances, varying by only 1-3\% when assumed 
$kT=4.5-7.5$ and $Z_{Fe} =10-30\%$. 
The central surface brightness is thus  $8.6 \times 10^{-13} \sb$. 

We estimated a gas mass for this cluster by converting the central
HRI surface brightness into a central electron density of 
$2.0 \pm ^{0.6}_{0.5} \times 10^{-2} h^{1/2}$ cm$^{-3}$ using the relation 
of Henry \& Henriksen (1986), modified to account for redshift effects
on surface brightness (Hughes \& Birkinshaw 1998). 
The gas mass is integrated out to a radius of 0.5 $h^{-1}$~Mpc by
assuming $\rho = 1.14 n_e m_H [1+ (r/r_{c})^2]^{-3\beta/2}$, 
with values of $r_{c}$ and $\beta$ from our fit to the HRI 
observations. The value 1.14 is the mass 
per free electron in a fully-ionized, 
primordial hydrogen/helium plasma in units of the proton mass, 
$(1.+4A_{He})/(1.+2A_{He})$
where $A_{He}$, the number fraction of He/H is 0.08.
The gas mass interior to $0.5\, h^{-1}$~Mpc found in this way is 
$0.9-1.2 \pm^{0.2}_{0.3} \times 10^{13} h^{-5/2} M_\odot$, for $q_0=0.5$ 
and $0.1$ respectively.
The main uncertainty in the central electron density and the gas mass 
inside 0.5~h$^{-1}$ Mpc arises from our fit to the HRI data. The 
uncertainties reflect the 90\% statistical uncertainties 
in $\beta$, $r_{c}$, and the appropriate fit normalization.

The total mass interior to $0.5 \, h^{-1}$~Mpc can be estimated if we assume 
that the ICM is nearly isothermal and in hydrostatic equilibrium within
the cluster gravitational potential. The total mass inside a radius of 
$0.5 \, h^{-1}$ Mpc would then be $2.1 \times 10^{14} \pm ^{1.5}_{0.8} h^{-1} 
M_\odot$ (relatively insensitive to $q_0$).  
The main sources of uncertainty are in the $\beta$ value for 
the gas distribution ($\beta$ is proportional to the total mass under our
assumptions) 
and the absolute temperature.  The quoted uncertainty does not reflect 
the systematic uncertainty in the assumption of hydrostatic equilibrium, 
which, according to current numerical simulations (e.g., Navarro, Frenk \& 
White 1995; Evrard 1997; Roettiger, Stone \& Mushotzky 1997; 
Roettiger, Burns, \& Loken 1996), can be considerable 
($\sim 20-50\%$) but less than the statistical uncertainities of our data.
The minimum mass estimate from weak lensing in this cluster is 
$2.45 \pm 0.8 \times 10^{14} h^{-1} M_\odot$ interior to 0.5~$h^{-1}$ Mpc 
(Clowe \etal 1998), consistent with the X-ray derived mass inside
the same radius.

The gas fraction within $0.5 \, h^{-1}$~Mpc is therefore $0.02-0.08 \, h^{-3/2}$, 
but this ratio could be as high as $0.10 \, h^{-3/2}$ if both $\beta$ 
and the actual virial temperature are at the low end of their
respective 90\% uncertainty 
ranges. Furthermore, according to the simulations mentioned in the previous
paragraph, cluster masses
derived from X-ray temperatures can exceed their actual masses 
by up to $20\%$, so the true gas fraction could be up to 20\%  
higher ($f_g \sim 0.12 \, h^{-3/2}$)  
if the gravitational mass is overestimated.  
These observations, therefore, do not necessarily imply that the gas fraction 
in this cluster is any different from the gas fractions typical of 
lower-redshift clusters.  

In earlier work, Donahue (1996) reported that the cluster MS0451.6-0305
showed evidence for a lower gas fraction than is typical of local clusters.
A re-analysis of the HRI data for the cluster MS0451.6-0305 (z=0.54) 
using the techniques described here
give a central surface brightness of $0.029$ HRI counts s$^{-1}$ 
arcmin$^{-2}$ and a best-fit $\beta = 0.6_{-0.05}^{+0.2}$ and 
$r_{core}= 0.5_{-0.05}^{+0.25}$ arcminutes (90\% uncertainties for
two interesting parameters.) The corresponding central electron density 
is $n_e = 0.012 ^{+0.001}_{-0.002} h^{1/2}$ cm$^{-3}$, and the 
gas mass inside 0.5 $h^{-1}$ Mpc is  
$2.4-2.7 \pm 0.2 \times 10^{13} h^{-5/2} M_\odot$, for $q_0=0.5$ 
and $0.1$ respectively. This estimate corrects the value in Donahue (1996)
which omitted a $(1+z)^{3/2}$ factor for the central density. The
gas fraction for MS0451.6-0305 is thus revised upward to span
$0.048 - 0.096 h^{-3/2}$, spanning the 90\% ranges of allowed $\beta$,
core radius, and $kT=10.9 \pm 1.2$ (Donahue 1996; this work). Thus we find no 
evidence that the gas fraction has changed in cluster cores from either
of the clusters MS1137.5+6625 or MS0451.6-0305. (The HRI data for 
cluster MS1054.4$-$0321 does not allow for straightforward 
conversion of the measured emissivity to gas mass since the X-ray 
emission profile shows internal structure and the cluster 
may not be hydrodynamically relaxed (Donahue \etal 1998).)

The cooling time within the core of MS1137.5+6625, estimated from the 
ratio $(5kT/2) / (\epsilon_{ff}(T)/n_e)$, where $\epsilon_{ff}(T)$ is the
free-free emissivity of hydrogen gas at temperature $T$, is somewhat 
shorter than the Hubble time at $z = 0.78$.  We assumed 
$\epsilon_{ff}(T) = 1.4 \times 10^{-27} g_{ff} [1.14 n_e^2] T^{0.5}$
erg cm$^{-3}$ s$^{-1}$, where $g_{ff} \sim 1.25$, $T$ is the electron
temperature in units of Kelvins, $n_e$ is the electron density, and
the factor 1.14 comes from the assumption that the gas is fully ionized. 
The cooling rate is estimated by computing the amount of gas inside the
radius at which the cooling time equals the Hubble time at the redshift
of the cluster, and dividing by the Hubble time at the cluster redshift.
For a range of cosmological
assumptions ($H_0=50-75$ km s$^{-1}$ Mpc$^{-1}$, $q_0=0.0-0.5$), the
derived cooling rate lies between 20 and 400 M$_\odot$ yr$^{-1}$, where 
the slower rates correspond to $q_0=0.5$. The cooling rates estimated
in this way are far more sensitive to $q_0$ than to $H_0$. For 
$H_0=75$ km s$^{-1}$ Mpc$^{-1}$ and $q_0=0.1$, the inferred cooling rate is
130 M$_\odot$ yr$^{-1}$.

\subsection{Keck-II Observations}

Multi-object spectroscopy of galaxies in the cluster MS1137.5+6625
was obtained with LRIS (Oke \etal 1995) on the Keck-II telescope 
on Feb. 17, 1998. We used a 1.5$^{\prime\prime}$ slit
width and a 300 line/mm grating at 5000 \AA, with the GG495 filter.
The wavelength scale was 2.47 \AA/pix; the spatial scale was
0.215$^{\prime\prime}$/pixel.
The seeing was 0.8$^{\prime\prime}$ - 1.0$^{\prime\prime}$. Three masks 
with total exposure times of 6300, 6900, and 7200 seconds were used to obtain spectra.

Twenty-three galaxies in MS1137.5+6625 have concordant redshifts
(Table~\ref{galaxies}). One galaxy
was discarded by the three-sigma clipping code used to compute the
velocity dispersion (Danese, De Zotti, \& Di Tullio 1980), for 22 
cluster members in total. 
The dispersion for the 22 galaxies is $884 \pm ^{185}_{124}$ km/sec 
(1-sigma errors), with a mean redshift of $z=0.7842 \pm 0.0003$. A
similar result was obtained using ROSTAT (Beers, Flynn \& Gebhardt 1990).
The velocity dispersion corresponds to a cluster temperature of 
$kT \sim \mu m_p \sigma_{\rm 1D}^2 = 4.8\pm0.4$ keV 
(1-sigma errors). Although the velocities of the  
cluster galaxies may appear to be a little ``cooler'' than the cluster
gas, in fact the temperature and velocity dispersion are statistically
in agreement. The temperature and velocity dispersion of 
MS1137.5+6625 are completely
consistent with the observed relationship for lower redshift 
clusters (Mushotzky \& Scharf 1997) and for other $z=0.5-0.8$ EMSS
clusters (Donahue \etal 1998). The agreement of the X-ray temperature
and the velocity dispersion and the agreement of 
the X-ray mass and the lensing mass from Clowe \etal (1998) 
suggest that the cooling flow in
this cluster does not affect the emission-weighted temperature 
significantly.

\subsection{Iron Abundance at $z \sim 0.8$}

The iron line detection in MS1137.5+6625, if real, is the most distant 
iron line detected in the EMSS cluster sample.  The best-fit 
abundance is consistent with the abundances of low-redshift 
clusters; however, the uncertainty range is disappointingly
large.  In order to improve our constraints on ICM abundances
at $z \sim 0.8$, we can combine our ASCA data on MS1137.5+6625 with
our existing ASCA data on MS1054.4$-$0321, the other EMSS cluster at 
this redshift (Donahue \etal 1998).

We estimated a ``mean'' iron abundance of clusters at 
$z \sim 0.8$ by simultaneously fitting all of the ASCA data for 
MS1137.5+6625 and MS1054.4$-$0321. We allowed the temperatures 
of the two clusters 
to differ, but required their metal abundances to be equal. 
The Galactic absorption columns to each cluster and the redshifts 
were fixed at their individual values. The resulting iron abundance 
was $Z_{Fe} = 0.33 \pm^{0.26}_{0.23}$ for a formal 90\% confidence 
range for one interesting fit parameter ($\Delta \chi^2 = 2.70$.)

The mean iron abundance at $z \sim 0.8$ is statistically 
more significant than
the abundance determinations for the individual clusters and is
similar to that of clusters of galaxies at lower redshifts. This result implies
 that the metal enrichment of the cluster ICM,
presumably from supernovae erupting in cluster galaxies, must have
occurred before $z \sim 0.8$.

\section{High Redshift Complete Sample of Clusters of Galaxies}

A primary motivation for observing distant clusters with ASCA
has been to establish a temperature function for high-redshift clusters
that can be compared with cosmological models.  Now that we have
ASCA data in hand for almost all the EMSS clusters at $z > 0.5$, we
can construct a high-redshift temperature function that is
based on a complete sample of clusters (Table~\ref{sample}).  
The flux limit of our subsample
of the EMSS, set by ASCA's capabilities, is approximately 
$1.80\times10^{-13} \, \flux$ in the EMSS detection cell of 
$2.4 \times 2.4$ arcminutes, within the Einstein bandpass 
of 0.3-3.5 keV.  Some of the original clusters listed
in Henry \etal (1992) are revised in our updated sample. 
We have revised the redshift of cluster MS1241.5+1710 upward from
0.31 to 0.54, based on spectroscopy of the cD galaxy and one other
galaxy. The cluster candidate MS2053.7-0449 at $z=0.58$ drops out of 
the sample because short followup observations by the ROSAT HRI failed
to detect it. Its X-ray flux is thus likely to be below the flux limits of 
our sample. However, it was definitely detected by the Einstein 
IPC, and long ROSAT HRI and ASCA observations have been made 
(PI: Henry). 
The cluster candidate MS1610.4+6616 (Gioia \& Luppino 1994) 
is an X-ray point source and is excluded from our sample, since it is a
likely AGN.

The resulting EMSS sample of clusters of galaxies with redshifts
greater than 0.5 is listed in Table~\ref{sample}. The X-ray temperatures
for the cluster MS 1241.5+1710 were extracted and fit from archival 
ASCA data using the same procedure as described above. We also analyzed 
the complete suite of ASCA data for the MS0015.9+1609 cluster, 
excluding the SIS0 and SIS1 dataset of the second observation contained in the
HEASARC because they exhibited severe anomalies. This analysis 
differs somewhat from the Hughes \& 
Birkinshaw (1998) analysis in that our analysis included 6 out of the
8 ASCA datasets available as well as the ROSAT PSPC data, while 
Hughes \& Birkinshaw (1998) analyzed the PSPC data and the GIS data from
the Performance Verification stage of the ASCA mission only. However, the
mean temperature that we obtain ($8.7 ^{+1.8}_{-0.6}$ keV, 90\% confidence
limits) is statistically consistent with their temperature
limits of $7.55 ^{+0.72}_{-0.58}$ keV ($1-\sigma$ errors), and all 
other best fit spectral values (Fe/H and $N_H$) are
nearly identical. These fits are also consistent with the
results of Furuzawa \etal (1998) who obtain a best fit temperature
of $8.0\pm^{1.0}_{0.8}$ keV.  The detection cell
fluxes (Table~\ref{sample}) are from the Einstein IPC (0.3-3.5 keV).   
The total fluxes within a 5-6 arcmin aperture were measured from the
ASCA GIS3 observations of each cluster. (The SIS observations do not
in general provide a good estimate of the total flux because flux
is lost off the side of the main CCD or between the CCDs.) Luminosities
were computed within XSPEC (Arnaud 1996), 
assuming the appropriate cosmological
models. We could
recover, to within uncertainties, estimates of the Einstein detect cell
flux from the large aperture ASCA GIS fluxes and estimates about
the cluster structure (see Table~\ref{sample}). From the
GIS total fluxes, we
determined the bolometric X-ray luminosity for each cluster (by using
the {\em dummyrsp} capability in XSPEC). The original Einstein detection
fluxes were used to compute 
maximum volume $V_{\rm max}$ within which the X-ray emission from the cluster
would exceed the flux limits of the EMSS, assuming that all clusters
have an average core radius of $0.125h^{-1}$ Mpc (Table~\ref{sample}).

We derived maximum volumes for each of the clusters following the 
prescription of Henry \etal (1992), using the Einstein 
detection cell fluxes and 
assuming a mean core radius of $125 h^{-1}$ kpc and $\beta=2/3$. The assumed 
source parameters did not influence the maximum volumes significantly, 
so we simply assumed the same parameters for all of the clusters.
For most of the clusters, the maximum volume spanned the redshift
limits of the sample ($0.5 \leq z \leq 0.9$). The cumulative 
temperature function was then computed by summing 
$1/V_{\rm max}$ for each $T \geq T_i$.  

\begin{figure}
\plotone{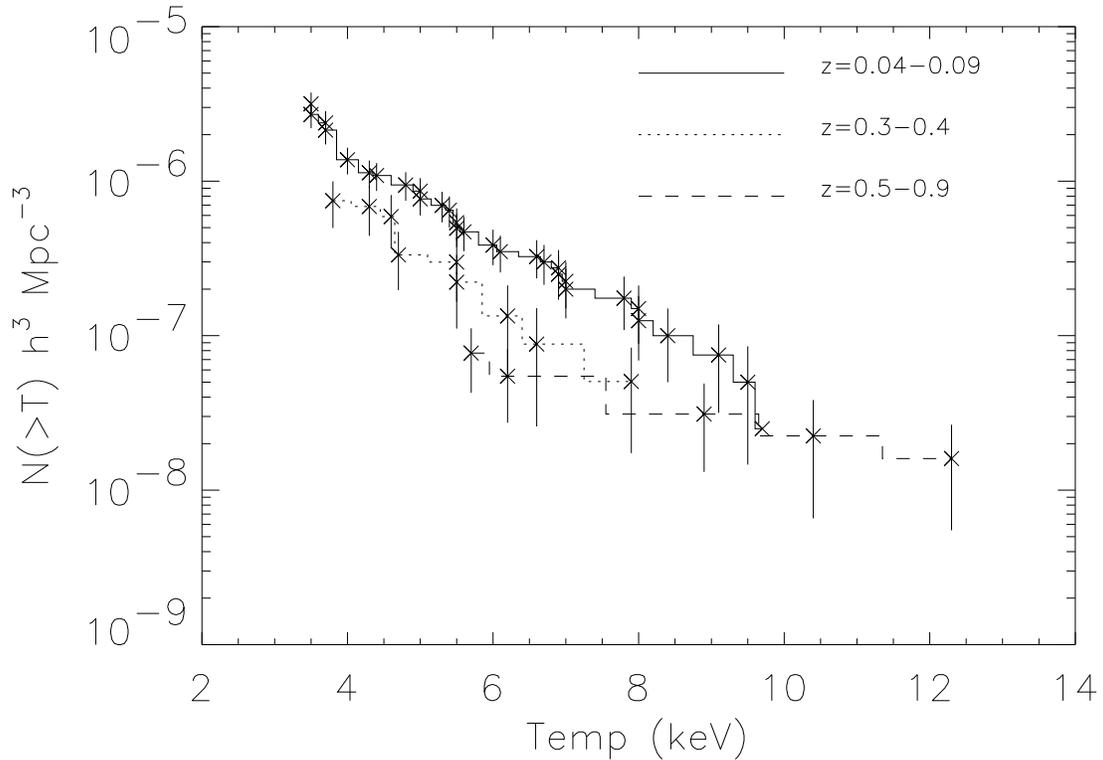}
\figcaption[ms1137_tf_q05.ps]{The temperature function at three different redshift ranges from
$z=0.04-0.09$ (data from Markevitch 1998) in solid line, $z=0.3-0.4$
(Henry 1997, sample revised as discussed in the text), dotted line, 
and $z=0.5-0.9$ (this paper) for $q_0=0.5$ in the dashed line. The 
temperature function changes slightly from one redshift range
to the next. $N(\geq T)$ is in the units of number of clusters per
cubic Mpc (Mpc$^{-3} h^3$) and temperature is plotted in
keV. The evolution in the temperature function is consistent with
$\Omega = 0.3-0.45$ (Donahue \& Voit 1999.) \label{tf}}
\end{figure}

The temperature function for these high-$z$ clusters differs only
modestly from the temperature functions for lower-redshift clusters.
Figure~\ref{tf} shows the temperature functions for clusters at
$z=0.04-0.09$ clusters (data from Markevitch 1998), $z=0.3-0.4$ 
(Henry 1997), and $z=0.5-0.9$ (this paper) for a $q_0 = 0.5$
cosmology.  The statistical review of these data and comparison to 
semi-analytic models are 
continued in our analysis paper (Donahue \& Voit 1999), which finds
that this amount of evolution in the temperature function is consistent
with $\Omega_M=0.45\pm0.1$ (open universe) or $\Omega_M=0.27\pm0.1$ 
(flat universe), with systematic uncertainties of an additional
$\pm0.1$ (see also Donahue 1999 for our preliminary report of this
work).
As we reported in Donahue \etal (1998), the presence of hot clusters
($> 8 \, {\rm keV}$) at $z > 0.5$ in the EMSS strongly suggests 
that $\Omega_M < 1$.

\section{Bolometric Luminosity-Temperature Evolution}

The relationship between the cluster X-ray luminosity and the
mean temperature of its ICM and the evolution of that relationship 
are relevant both for what we can learn   
about cluster ICM physics and for tests of cosmology from studying
the evolution of the cluster luminosity function. 

At low redshift, the X-ray luminosities of galaxy clusters are
related to their temperatures via a power law: $L_x \propto 
T_x^\alpha$ with $\alpha \sim 3$ (Edge \& Stewart 1991; 
David \etal 1993, Arnaud \& Evrard 1998).  
Not all clusters are isothermal
and the scatter in this relation can be reduced if one corrects 
for the cooler gas often found in cluster cores.  
Recent relations derived by excluding the cool regions 
(Markevitch 1998) or by accounting for them with models 
(Allen \& Fabian 1998) arrive at somewhat flatter power-law 
slopes.  All the cluster temperatures and
luminosities in our high-$z$ EMSS sample are consistent 
with these low-$z$ temperature-luminosity relations. Here we 
assess the significance of that consistency. 

The scaling of cluster luminosities with cluster temperatures
presumably depends on the thermal history of the ICM.
In general, the luminosity of a cluster should scale with the core
density ($n_c$), the core radius ($r_c$), and the temperature: 
$L_X \propto n_c^2 r_c^3 T^{1/2}$.  
According to self-similar models of cluster formation in a
critical universe, the cluster temperature should scale as 
$T_X \propto n_c r_c^2$, and given $n_c \propto (1+z)^3$, one finds
$L_X \propto (1+z)^{3/2} T^2$ (Kaiser 1986).  The power-law
dependence of $L_X$ on $T_X$ in this relation is somewhat shallower
than observed.  One way to account for this discrepancy is to suppose 
that the ICM of each cluster undergoes a similar pre-heating 
event that fixes the minimum specific entropy at a value common
to all (Kaiser 1991; Evrard \& Henry 1991).  
Then the core density is determined by the ICM 
temperature ($n_c \propto T^{3/2}$ for an ideal gas), and the 
luminosity should scale as $L_X \propto T_X^{11/4}$, with redshift
evolution depending on how and when the cluster gas heats or 
cools (e.g., Bower 1997).

Furthermore, the scaling of cluster luminosities with cluster 
temperatures is an important step in the exercise of connecting
the observed X-ray luminosity function of clusters at different
redshifts with predictions of the evolution of the cluster mass
function. If the evolution of the cluster $L_X - T_X$ relation is strongly
positive ($L_x \propto T_x^3 (1+z)^A$, where $A \sim 2$), then the 
lack of evolution in the cluster
luminosity function, at least at $L_x < 3 \times 10^{44}~ \lum$ 
(Rosati \etal 1998; 
Ebeling \etal 1998; Sadat, Blanchard,
\& Oukbir 1998; Mathiesen \& Evrard 1998), is 
remotely consistent with the strong evolution of the mass function 
predicted by $\Omega_M=1$ models (Borgani \etal 1999). This consistency
arises because
the evolution in the mass function may be counteracted by evolution in
the $L_X-T_X$ relation (see also Oukbir \& Blanchard 1992 for the
EMSS sample). 

In order to constrain the redshift evolution of the $L_X$--$T_X$
relation with observations, we have compiled data on $z>0.1$ cluster 
bolometric luminosities, temperatures, and redshifts from  
Mushotzky \& Scharf (1997), Donahue (1996), Donahue \etal (1998), 
and Henry (1997).  We have added cluster AX2019 (Hattori \etal 1997) 
to this sample and have revised the redshift, luminosity, and 
temperature of the cluster MS1241.5+1710 as described above.

Since we cannot correct our high-redshift
temperatures for deviations from isothermality with our current data, 
we have chosen 
to compare mean temperatures of clusters at low redshift with 
mean temperatures of clusters at high redshift and to quantify  
the higher intrinsic dispersion incurred.  
For simplicity, we compare the data with an $L_X$--$T_X$
relation assuming a power-law redshift dependence: 
$L_x \propto T_x^\alpha (1+z)^A$ (Borgani \etal 1999). 
Formally, the observational uncertainties in the temperature measurements
are often significantly smaller than the scatter in the relation, demonstrating
that there are physical effects that are not taken into account by
a simple relation, such as cooling flows (see Allen \& Fabian (1998);
Markevitch (1998) for how these effects might be handled in X-ray data with
sufficient resolution and signal to noise.)

\begin{figure}
\plotone{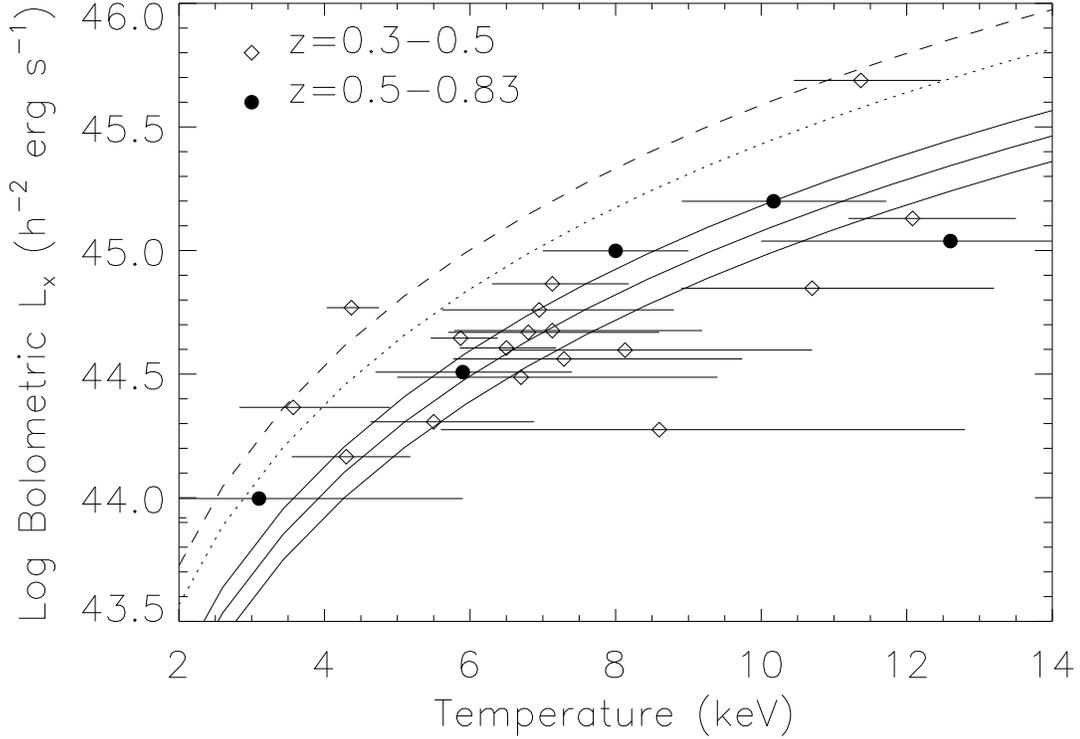}
\figcaption[lt.ps]{Bolometric luminosities and temperatures for $z>0.3$ 
EMSS
clusters and other clusters from the literature are plotted here. The
filled circles are clusters with $z=0.5-0.9$, and the open diamonds
are clusters with $z=0.3-0.5$. The solid lines indicate the low-redshift
bolometric luminosity temperature relation from Markevitch (1998) where
$L_x \propto T_x^{2.64}$, and a dispersion in log $L_x$ of 0.103. The 
luminosities in this plot were computed assuming $q_0=0.5$. The dotted
and dashed lines indicate where the low-redshift relation evolves if
$A=2$ and $z=0.3$ and $z=0.8$ respectively. Note that the Markevitch
relation is for clusters whose central cooler regions have been removed
from the data, while the $z>0.3$ data are uncorrected. Corrected cluster
temperatures would tend to move to the right in this plot. The
two most serious outliers are $z\sim 0.4$ clusters from the Mushotzky
\& Scharf (1997) sample,   
RXJ1347.5-1145 ($kT \sim 11.4$ keV) and MS2137.3-2353
($kT \sim 4.4$ keV). They are 
both massive cooling flow clusters with unusually high luminosities
for their temperatures (Allen \& Fabian 1998). 
\label{lt} }
\end{figure}

For illustrative purposes, we plot the $z=0.3-0.5$ and the $z=0.5-0.9$
bolometric luminosity and temperature 
data in Figure~\ref{lt}, along with the Markevitch bolometric 
$L_x-T_x$ fit to the relation, with $1\sigma$ dispersion 
to  corrected temperatures and luminosities
of low-redshift clusters ($L_x = 3.11 \times 10^{44} h^{-2} \lum T_6^{2.64}$
where $\sigma_{log~L}=0.103$). We note that the uncorrected high-redshift
points cluster around the low-redshift relation within the dispersion.
In Figure~\ref{lt}, the normalization of the low-redshift Markevitch
relation has been evolved appropriate to $A=2$ for two epochs, 
$z=0.3$ and $z=0.8$. This amount of evolution is ruled out by the data.

To test the significance to which evolution of the $L_x-T_x$ relation 
could be constrained, 
the procedure was to obtain a linear regression fit
of the expression 
$\log T_{\rm fit} = \log T_0 + (1/\alpha) \log L_x - (A/\alpha) \log (1+z)$
to the redshift, luminosity, and temperature data. 
We weighted each point by defining an effective standard
deviation $\sigma_{\rm eff} = 0.6 (T_{h} - T_{l})/2$,
where $T_h$ and $T_l$ are the upper and lower bounds on
the 90\% confidence interval of the temperature.  
This value for $\sigma_{\rm eff}$ roughly reproduces the width 
of the 68\% confidence interval normally associated with $1 \sigma$. 
The David \etal (1993) cluster temperature catalog provides 
both 68\% and 90\% confidence
limits. The mean ratio between the 68\% intervals and the corresponding 
90\% intervals is $0.60\pm 0.06$ for $kT > 6$ keV. 

To check this estimate, the actual probability distributions for each of the
measurements were derived for the 5 EMSS clusters in our highest
redshift sample. We then fit the probability distributions to Gaussians. The
results of this exercise demonstrated that the estimate of the size
of the 68\% uncertainties from the 90\%
uncertainties as obtained from the literature 
is reasonable. However,  we also note in the cases where
the error bars are uneven in size (the upper error bar is larger than the
lower error bar), the mean temperature from the Gaussian fit is higher
than the actual best-fit temperature, and the errors from the Gaussian
fit are approximately the mean of the two original errors (Table~\ref{chisq-fit}.) 
 
Formally, the best fitting $L_x - T_x$ relation is not a 
good fit, owing to the significant intrinsic scatter in the relationship between
luminosity and temperature. Intrinsic
scatter can be caused by inclusion of ``cooling-flow'' emission from the core
of the galaxy and other processes that affect the luminosity of the 
cluster, 
which is very sensitive to the electron density in the ICM. 
To ensure that our $\chi^2$ values reflected the actual scatter in the
$L_x - T_x$ relation, we added a constant intrinsic scatter term to 
each measurement's standard deviation in quadrature, so that the
reduced $\chi^2=1$ for the best fit.
 In this case, the intrinsic scatter of the $L-T$ relationship
in the temperature of clusters with 
$kT > 3$ keV and $z=0.1-0.9$ is $\sim0.08$ in units of log keV. The intrinsic
scatter in log T for these clusters is nearly identical to that of 
low redshift clusters. Markevitch (1998) finds an intrinsic scatter
of $\sigma_{{\rm log~} T} \sim0.09$ for uncorrected temperature data. 
Mushotzky \& Scharf (1997) also noted that the intrinsic variance in 
the $L-T$ did not change with redshift in their analysis of a subset
of the clusters analyzed here, with $z=0.14-0.55$.

\begin{figure}
\plotone{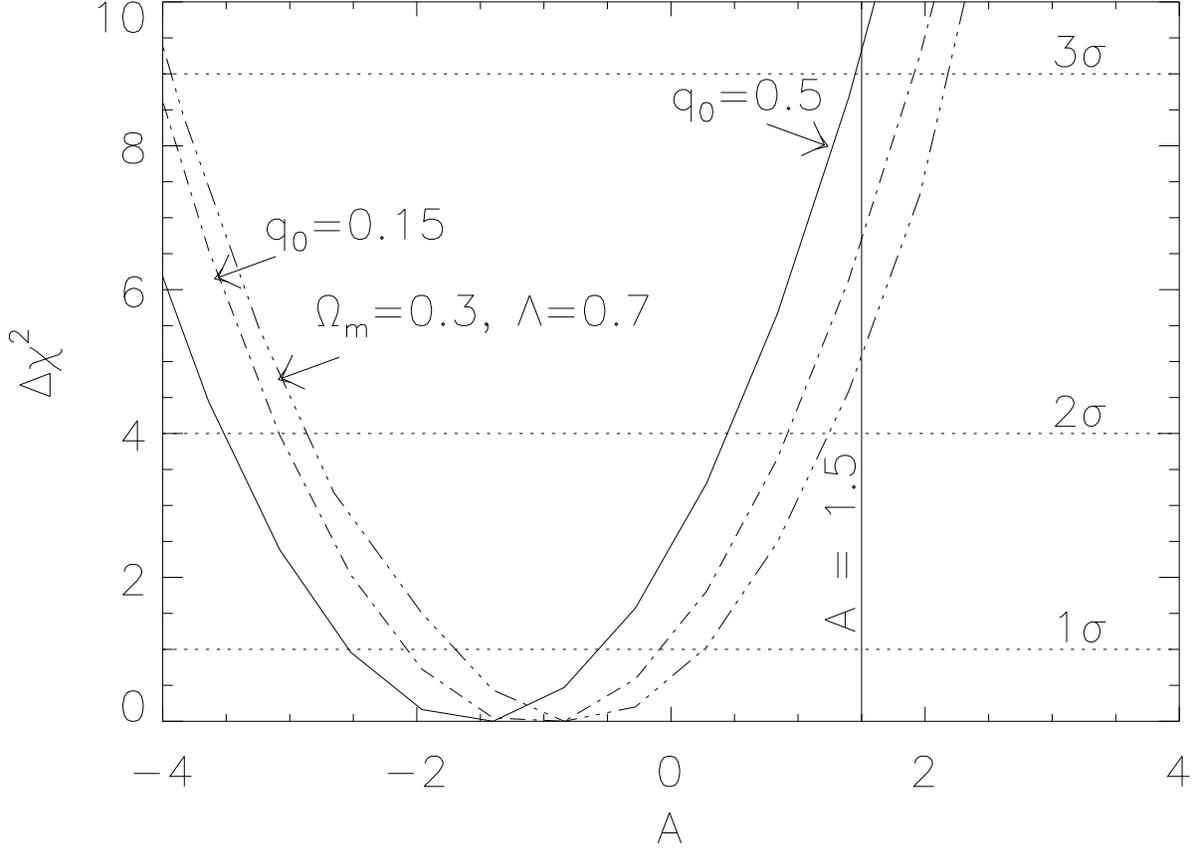}
\figcaption[qo_eq_all.eps]{The change in $\chi^2$ for a fit to a single interesting
parameter, $A$, in which the luminosities and temperatures of
a sample of moderate redshift ($z=0.1-0.9$) clusters is fit to the relation 
$L_x \propto T_x^{\alpha} (1+z)^{A}$, including an intrinsic
dispersion. (See text for details.) Luminosities were
computed assuming  $q_0=0.5$ (solid line), 
$q_0=0.15$ (dashed line), and a flat universe where
$\Omega_m=0.3$ and $\Lambda=0.7$ (dot-dash line). 
One, two and three sigma levels are marked horizontally
across the plot. A model
where $A=3/2$ is ruled out at $>3$ sigma for the $q_0=0.5$.
\label{qo5}}
\end{figure}

Because the observed luminosities of clusters at these
moderate redshifts depend on the assumed cosmology,
we fit the $L_X$--$T_X$ relation for three different cosmologies. Our
results for the best fit $A$ vs. $\Delta \chi^2$ are displayed in Figure~\ref{qo5}. 
The most relevant cosmology for $\Omega_{M}=1$ models is the
flat $q_0=0.5$ assumption. 
The $\Delta \chi^2$ distribution for $A$ in a critical ($q_0 = 0.5$) 
cosmology, shown in Figure~\ref{qo5}, 
demonstrates that self-similar evolution
($A = 1.5$) is a poor fit to the data, even when intrinsic dispersion
is accounted for.  The best-fit value of 
$A$ is actually slightly negative and $\Delta \chi^2 = 9.5$
for $A = 1.5$.  Our results are statistically consistent with those 
obtained by Reichart, Castander, \& Nichol (1999), who assumed all of
the EMSS clusters had a cluster temperature of 6 keV and who fit only
the $z\leq 0.5$ clusters.

The absence of $L_X$--$T_X$ evolution in our sample poses 
serious difficulties for cluster evolution in a critical
universe. The reason for this difficulty is that in order to 
account for the lack of evolution observed in the cluster luminosity
function out to $z\sim0.8$ (Rosati \etal 1998), models in which  
$\Omega_M=1$ must include sufficient luminosity evolution to mask
the dramatic number evolution in the mass function  
that occurs if $\Omega_M=1$ (Viana \&
Liddle 1996; Eke, Cole \& Frenk 1996). 
The power-law index of $A \approx 2-3$ which reproduces the lack of evolution
of the observed cluster
luminosity function at the outer limits of statistical and systematic  
uncertainties (Borgani \etal 1999)
results in an extremely poor fit to our high redshift $L_X$--$T_X$ data for
any  $q_0$. 

\section{Conclusions}

We have presented X-ray observations from the ASCA and ROSAT satellites, and
new Keck-II galaxy redshifts for the EMSS cluster MS1137.5+6625 at $z=0.78$. This
cluster is the last cluster in our complete sample of high-redshift
EMSS clusters of galaxies to be observed with ASCA. 
X-ray spectra from ASCA constrain the
temperature of the intracluster plasma to be $kT = $~ \asca_kT 
keV. The total X-ray luminosity within a 5$^\prime$ radius GIS aperture is
$1.9-2.8 \times 10^{44}\,h^{-2}$ erg s$^{-1}$ in the 2-10 keV band
in the rest frame of the cluster ($q_0=0.0-0.5$). 
The luminosity and temperature are
consistent with the empirical relation between X-ray
luminosity and redshift for low-redshift clusters of galaxies.

The velocity dispersion of 22 member galaxies measured with the Keck-II
telescope and LRIS is $884 \pm^{185}_{124}$ km/sec, which is consistent
with the value of the ICM temperature. This consistency implies that 
the thermal properties of the X-ray gas and the dynamics of the galaxies are
both governed by the gravitational potential of the cluster. Clowe \etal (1998)
report a  weak lensing mass of $2.45 \pm 0.8 \times 10^{14}\,h^{-1}\, M_\odot$ interior
to 0.5 h$^{-1}$ Mpc. 
This is consistent with the isothermal mass 
estimated from the cluster temperature and the best-fit $\beta=0.7$ from
the HRI observations of $2.1 \times 10^{14}\,h^{-1}\,M_\odot$. Therefore the
cluster temperature, the galaxy velocity dispersion, and the 
weak lensing mass, all independent measures of the cluster gravitational
mass, are all self-consistent at a radius of 0.5~$h^{-1}$~Mpc.

We report a mean iron abundance for $z\sim 0.8$ clusters
of galaxies that is consistent with that of low redshift clusters
of galaxies by simultaneously fitting all of the ASCA spectra for
MS1054.4$-$0321 and MS1137.5+6625. The mean abundance is $0.33$ solar with a formal
one-dimensional 90\% uncertainty range of 0.10--0.59. 
This measurement is one of the highest-redshift detections
of intracluster iron.

We have estimated a gas fraction that is consistent with being
the same as that of clusters at low redshift. Our main uncertainties
are primarily in constraining the gas distribution of the cluster and
secondarily in constraining the cluster temperature. The cluster may
have a modest cooling flow of $\sim 20-400$ M$_\odot$ yr$^{-1}$.
In summary, the velocity dispersion, temperature, gas fraction, and
iron abundance of MS1137.5+6625 are all similar to those properties
in lower redshift clusters of similar luminosity. 

The X-ray luminosity-temperature relation for clusters appears to evolve
little out to $z \sim 0.8$. We supplemented the Mushotzky \& Scharf 
(1997) $L_x-T_x$ dataset with the two $z=0.8$ EMSS clusters, MS1054.4$-$0321 
and MS1137.5+6625, the $z=0.9$ cluster discovered by Hattori 
\etal (1997), the Henry $z=0.3-0.4$ clusters, and one of the Henry clusters 
revised to a higher redshift in order to constrain the parameter $A$, 
where $L_x \propto T_x^{\alpha} (1+z)^A$. We exclude $A=3/2$ at 
greater than $3\sigma$ confidence for $q_0=0.5$. Values of $A=2-3$, required
to explain why the X-ray luminosity function does not appear to 
change out to $z\sim0.8$ if $\Omega_m=1$, are strongly ruled out 
by this data. 

We present a cluster temperature function for $z=0.5-0.9$ based on 
a complete sample of 5 EMSS clusters.  A companion paper (Donahue \&
Voit 1999) compares 
this high-redshift temperature function to cosmological models which 
predict how the cluster temperature function should evolve.

\acknowledgements
MD acknowledges partial support from a NASA grants NAG5-3257, NAG5-3208, and
NAG5-6236. CAS is completely supported by NASA LTSA grant NAG5-3257. 
IMG and CRM acknowledge partial financial 
support from NSF grant AST95-00515 and NASA STSCI grant GO-05987.02-94A, 
and from CNR-ASI grants. JPH's research on clusters of galaxies is
partially supported by NASA LTSA Grant NAG5-3432. This paper has made use
of data obtained through the High Energy Astrophysics Science Archive
Research Center Online Service, provided by the NASA/Goddard Space Flight
Center.

\newpage

\begin{figure}
\plotone{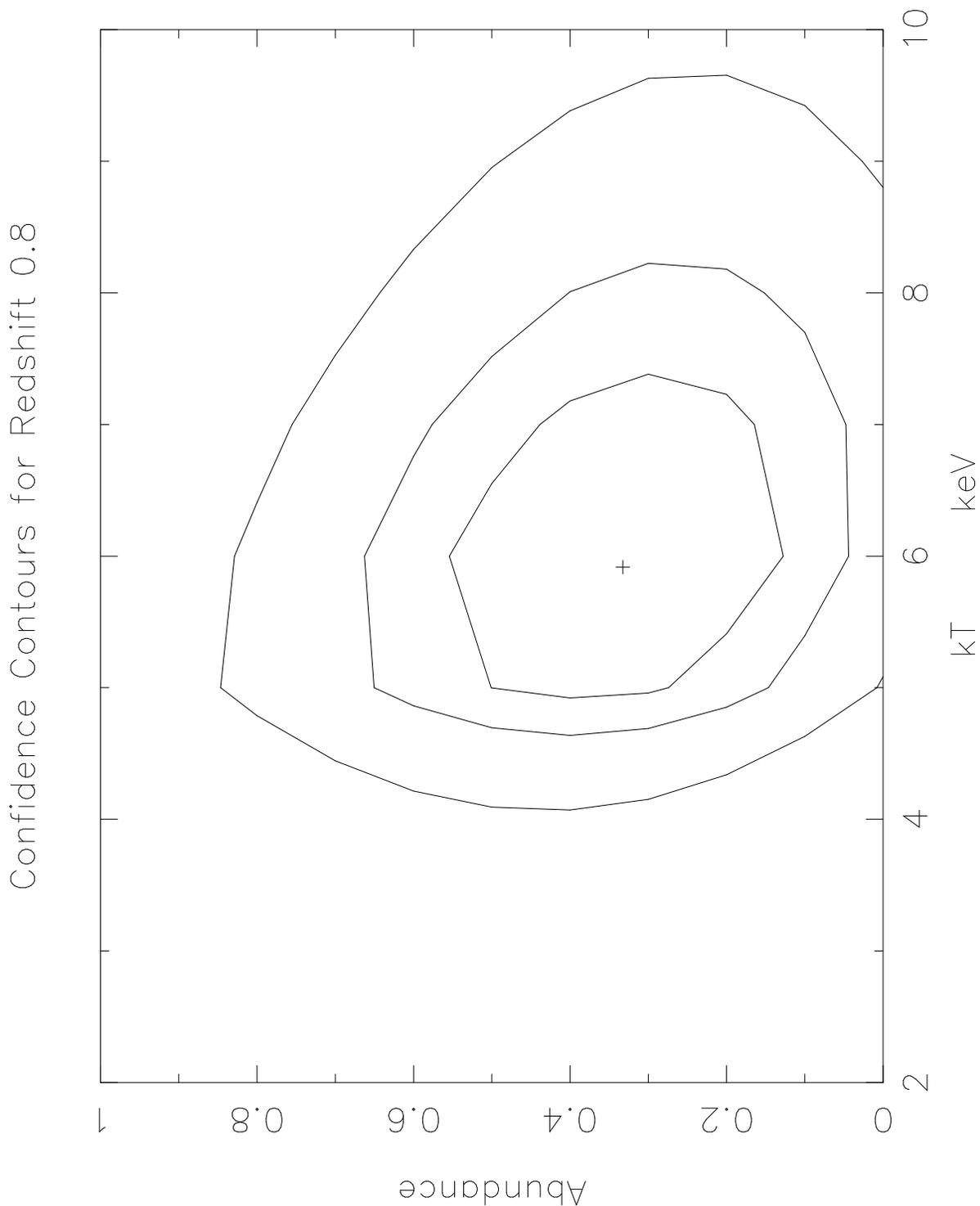}
\figcaption[ms1137_fe_all.eps]{We plot the result of a simultaneous fit of two
$z=0.8$ cluster iron abundances. We allowed the fluxes and the 
temperatures of the two clusters to be different but forced the
iron abundance to be the same. We plot the two-dimensional 
$\chi^2$ contours at 68.3\%, 90\% and 99\% confidence levels 
($\Delta \chi^2 = 2.30$, 4.61 and 9.21) for two interesting
parameters for the mean cluster 
iron abundance in units of the solar abundance 
and temperature of MS1137.5+6625 in keV.   \label{contour}}
\end{figure}

\newpage
\newpage

\begin{table}
\caption{ASCA Observations \label{asca}}

\begin{tabular}{lcc} \tableline \tableline
Detector & Good Exposure & Count Rate \\
    & (seconds) & ($10^{-3}$ cts/s) \\ \tableline
SIS0 & 71,377 & $6.7\pm0.4 $ \\
SIS1 & 70,528 & $4.5\pm0.3 $ \\
GIS2-spectrum & 66,432 & $3.7\pm0.4 $ \\
GIS3-spectrum & 66,494 & $4.4\pm0.4 $ \\ 
GIS2-5 arcmin & 66,432 & $4.5\pm0.6 $ \\
GIS3-5 arcmin & 66,494 & $6.1\pm0.7 $ \\ \tableline
\end{tabular}
\end{table}

\begin{table}
\caption{HRI Observations \label{hri}}
\begin{tabular}{llr} \tableline \tableline
Dataset ID  & Observation Dates & Exposure Time \\ \tableline
RH800662N00 & May 19-25, 1995 & 3,563 seconds \\
RH800784N00 & Oct 12-Dec 4, 1995 & 32,202 seconds \\
RH800784A01 & May 6-30, 1996  & 64,269 seconds \\ \tableline
\end{tabular}
\end{table}

\begin{table}
\caption{Redshift Catalog for MS1137.5+6625 \label{galaxies}}

\begin{tabular}{llrr} \tableline \tableline
RA (J2000) & DEC (J2000) & ID & Redshift \\
11 40 22.3 & +66 08 15  &    1  & $0.7844\pm0.0010$ \\
11 40 22.0 & +66 08 12  &    2  & $0.7903$\tablenotemark{a} \\
11 40 22.6 & +66 08 17  &    3  & $0.7844\pm0.0006$  \\
11 40 22.0 & +66 08 19  &    4  & $0.7927\pm0.0007$ \\
11 40 21.4 & +66 08 20  &    5  & $0.7826\pm0.0009$ \\
11 40 17.7 & +66 08 23  &    6  & $0.7808$\tablenotemark{a}\\
11 40 16.9 & +66 08 25  &    7  & $0.7909\pm0.0007$ \\
11 40 17.2 & +66 08 29  &    8  & $0.7916\pm0.0007$ \\
11 40 27.4 & +66 08 23  &    9  & $0.7761\pm0.0011$ \\
11 40 27.0 & +66 08 09  &   10  & $0.7760\pm0.0015$ \\
11 40 29.3 & +66 08 02  &   11  & $0.7831\pm 0.0001$\\
11 40 24.5 & +66 07 26  &   12  & $0.7823\pm0.0013$ \\
11 40 14.8 & +66 07 32  &   13  & $0.7824\pm0.0009$ \\
11 40 30.8 & +66 09 03  &   14  & $0.7900\pm 0.0011$ \\
11 40 15.1 & +66 09 22  &   15  & $0.7773\pm0.0008$ \\
11 40 17.4 & +66 09 14  &   16  & $0.7909\pm 0.0011$ \\
11 40 42.8 & +66 10 19  &   17  & $0.7814\pm0.0010$ \\
11 40 40.4 & +66 07 52  &   18  & $0.7889\pm0.0003$ \\
11 40 37.1 & +66 07 25  &   19  & $0.7902$ \tablenotemark{b}\\
11 40 10.3 & +66 07 18  &   20  & $0.7898\pm0.0008$ \\
11 40 05.9 & +66 08 05  &   21 &  $0.7790$\tablenotemark{a}\\
11 39 58.9 & +66 08 00  &   22  &  $0.7792\pm0.0007$\\
11 40 06.0 & +66 08 18  &   23  &  $0.7714$\tablenotemark{c}\\
 \tableline
\end{tabular}
\tablenotetext{a}{CaII break only.}
\tablenotetext{b}{[OII] emission line only.}
\tablenotetext{c}{[OII] only, clipped by 3-sigma code}
\end{table}

\begin{table}
\caption{Summary of EMSS Cluster X-ray Parameters \label{sample}}
\begin{tabular}{lccccc} \tableline\tableline

Cluster & MS0451-03 & MS1241+17 &MS0015+16 & MS1137+66 & MS1054-03 \\
Redshift & 0.539    & 0.540     & 0.5455    & 0.785     & 0.828     \\
kT (keV) &  $10.9 \pm 1.2$\tablenotemark{a} & $6.4 \pm^{1.8}_{1.1}$ & $9.9 \pm^{1.1}_{1.0}$ 
			& \asca_kT & $12.3 \pm^{3.7}_{2.1}$ \\
GIS aperture (arcmin) & 6.0 & 5.0 & 6.0\tablenotemark{h} & 5.0 & 5.0 \\
$F_x$ (GIS, 0.3-3.5 keV obs) \tablenotemark{b} & $19.0\pm0.4$ & $7.1\pm0.4$ & $21.0\pm0.4$ 
			& $2.7\pm0.3$ & $5.6\pm0.3$ \\
$F_x$ (GIS, 2-10 keV obs)\tablenotemark{b}     & $22.0\pm0.5$ & $6.2\pm0.3$ & $23.0\pm0.4$ 
			& $1.8\pm0.2$ & $6.1\pm0.3$ \\
$L_{44}$ (2-10 keV rest)\tablenotemark{c}      & $10.6\pm0.2$ & $3.4\pm0.2$ & $12.0\pm0.2$ 
			& $2.8\pm0.3$ & $8.4\pm0.5$ \\
$L_{44}$ (bolometric)\tablenotemark{c}         & $21.0\pm0.4$ & $6.1\pm0.3$ & $22.0\pm0.4$ 
			& $4.8\pm0.5$ & $16.6\pm0.9$ \\
$F_{x,det}$ (Einstein, 0.3-3.5 keV)\tablenotemark{d} & $9.54\pm1.71$   & $4.23\pm1.08$  & $7.06\pm0.79$ 
			& $1.89\pm0.36$ & $2.11\pm0.25$ \\
$F_{x,det}$ (GIS, 0.3-3.5 keV)\tablenotemark{e} & $12.4\pm1.2$ & $4.31\pm0.5$  & $12.8\pm0.3$ 
			& $2.46\pm0.3$ & $2.19\pm0.25$ \\
core radius (arcmin)\tablenotemark{f}& 0.5  & 0.68 & 0.74 & 0.25 & 0.9 \\
Vol, $q_0=0.5$\tablenotemark{g}& 15.58  & 4.24  & 11.60 & 4.45  &  6.25 \\
Vol, $q_0=0.15$\tablenotemark{g}& 21.97 & 5.73  & 16.26 & 6.77  &  9.67 \\
Vol,  $\Omega_M=0.3$,$\Lambda=0.7$\tablenotemark{g} & 34.12 & 8.59  & 25.91 & 10.78 & 15.55  \\
\end{tabular}

\tablenotetext{a}{The temperature for MS0451-03 is slightly revised from 
Donahue 1996; the spectra were
refit using the most recent versions of the SIS and GIS response matrices for ASCA.}
\tablenotetext{b}{X-ray fluxes are all quoted in units $10^{-13} \flux$. Uncertainties
reflect statistical uncertainties, not calibration uncertainties.}
\tablenotetext{c}{X-ray luminosities are in units $10^{44} h^{-2}$ erg s$^{-1}$, $q_0=0.0$.
$L_x$ is the X-ray luminosity estimated from the ASCA
GIS3 observation within the GIS aperture radius listed in this table.}

\tablenotetext{d}{ X-ray flux is the Einstein IPC flux measured in the detection cell 
in units of $10^{-13} \flux$.}
\tablenotetext{e}{ $F_x$ in central 2.4' by 2.4' aperture, 
as {\em estimated} from ASCA GIS flux
rates and $r_{core}$ from this table, with $\beta=0.67$. Uncertainties are estimated from counting statistics and
a 10\% systematic uncertainty.}
\tablenotetext{f}{Core radii, as estimated from 
ROSAT HRI data. Core radii values are only general estimates for MS1054 and MS1241. MS1054 is too
irregular and the HRI data for MS1241 do not permit simultaneous constraints of both
$\beta$ and $r_{core}$. The core radii of 
MS1137 and MS0451 are from this paper. MS0015's (CL0016) core radius is 
from Hughes \& Birkinshaw (1998).}
\tablenotetext{g}{Detection volumes are in units $10^7$ Mpc $^3h^{-3}$. Survey volumes were computed
from the Einstein detect cell fluxes, assuming a mean $r_{core}=0.125h^{-1}$ Mpc and $\beta=2/3$. }
\tablenotetext{h}{The GIS aperture here for MS0015 includes flux from a nearby QSO. The QSO contributes
approximately 10\% of the total flux between 0.2-2.0 keV (Hughes \& Birkinshaw 1998). The PSPC 
flux for the cluster, conservatively excluding the QSO, is $\sim73\%$ that of the GIS. Our measured count
rates are very similar to HB1998, but the flux calibration of the GIS has changed since.} 
\end{table}

\begin{table}
\caption{Comparison of Actual Fit Temperatures and 1-$\sigma$ Errors to Mean Fit Temperatures
\label{chisq-fit}}
\begin{tabular}{lcc}\tableline \tableline
Cluster & Actual Fit & Mean Fit from Gaussian \\
        & (keV), 1-sigma &  (keV), 1-sigma \\ \tableline
MS1054 & $12.3^{+1.9}_{-1.3}$ & $12.6 \pm 1.6$ \\
MS1137 & $5.8^{+0.9}_{-0.7}$ & $6.0 \pm 0.9 $ \\
MS0451 & $11.0^{+0.9}_{-0.8}$ & $11.1 \pm 0.9$ \\
MS0016 & $9.9^{+0.7}_{-0.6}$ & $9.9 \pm 0.6$ \\
MS1241 & $6.4^{+0.8}_{-0.7}$ & $6.5 \pm 0.8$ \\ \tableline
\end{tabular}
\end{table}

\end{document}